\documentstyle[12pt,aaspp4,psfig]{article}
\lefthead{Masciadri et al.}
\righthead{The precession of the giant HH34 outflow: a possible
jet deceleration mechanism}
\begin{document}

\title{The precession of the giant HH34 outflow: a possible jet deceleration
mechanism}

\author{E. Masciadri\altaffilmark{1}$^{,}$\altaffilmark{2}, E. M. de Gouveia Dal Pino\altaffilmark{3}, A. C. Raga\altaffilmark{4}, A. Noriega-Crespo\altaffilmark{5}}

\altaffiltext{1}{Instituto de Astronom\'\i a,
UNAM, Apdo. Postal 70-264, 04510 M\'exico, D. F., M\'exico,
email: elena@astroscu.unam.mx}
\altaffiltext{2}{Max-Plank Institut f\"ur Astronomie,
K\"onigstuhl 17, D-69117 Heidelberg, Germany, email: masciadri@mpia.de} 
\altaffiltext{3}{Instituto Astronomico e Geofisico (IAG-USP),
Universidade de S\~ao Paulo, Cidade Universitaria,
R. do Matao, 1226, S\~ao Paulo, SP 05508-090, Brasil,
email: dalpino@astro.iag.usp.br}
\altaffiltext{4}{Instituto de Ciencias Nucleares,
UNAM, Apdo. Postal 70-543, 04510 M\'exico, D. F., M\'exico,
email: raga@astroscu.unam.mx}
\altaffiltext{5}{SIRTF Science Center, California Institute of Technology,
IPAC 100-22, Pasadena, CA 91125,
email: alberto@ipac.caltech.edu}

\begin{abstract}
The giant jets represent a fundamental trace of the
historical evolution of the outflow activity over timescales of
$\sim 10^{4}$ yr, i.e. a timescale comparable to the accretion time
of the outflow sources in their main protostellar phase. 
The study of such huge jets provides the
possibility of retrieving important elements related to the life
of the outflow sources. In this paper, we study the role of
precession (combined with jet velocity-variability and the 
resulting enhanced interaction with the
surrounding environment) as a deceleration mechanism for giant jets
using a numerical approach. This thesis was proposed for the first
time by Devine et al. (1997) but it could not be numerically explored
until now because it is intrinsically difficult to reproduce,
at the same time, the large range of scales from $\sim 100$~AU up
to a few parsecs. In the present paper, we obtain predictions of
H$\alpha$ intensity maps and position-velocity diagrams from 3D
simulations of the giant HH~34 jet (including an appropriate ejection
velocity time-variability and a precession of the outflow axis),
and we compare them with previously published observations of this object.
Our simulations represent a step forward from previous numerical studies of
HH objects, in that the use of a 7-level, binary adaptive grid has allowed
us to compute models which appropiately cover all relevant scales of
a giant jet, from the $\sim 100$~AU jet radius close to the source to 
the $\sim 1$~pc length of the outflow.
A good qualitative and quantitative agreement is found between the
model predictions and the observations, indicating that
a precession of the jet axis can indeed be the probable cause of the
deceleration of the giant jets. Moreover, we show that a critical
parameter for obtaining a better or worse agreement with the observations
is the ratio $\rho_{j}/\rho_{a}$ between the jet and the
environmental densities. The implications of this result in the context
of the current star formation models are discussed
\end{abstract}

\section{Introduction}
\label{intr}
Herbig-Haro (HH) objects are the optical manifestations of
outflows from young stellar objects (YSOs).
Following the discovery of jet-like structures in HH
objects (Dopita et al. 1982; Mundt \& Fried 1983), 
many of these HH jets were observed in the
Orion (Reipurth et al. 1986; Mundt et al. 1987; Reipurth 1989a, 
Reipurth 1989b) and Taurus (Mundt et al 1988) star formation
regions. These objects present a characteristic morphology of
aligned knots extending over $\sim 0.3$~pc.
It appears that the kinematics and morphologies of these jets
depend simultaneously on the time-dependent nature of the
outflow activity and on the interaction of the hypersonic flows
with the surrounding interstellar medium.

It has recently been discovered (Bally \& Devine 1994;
Reipurth et al. 1997; Devine et al. 1997) that a few HH jets
extend over distances of a few parsecs. For example, HH~111 shows
a total extent of $\approx 7.7$~pc, HH~34
of $\approx 3$~pc and HH~355 a total extent of $\approx 1.55$~pc.

Apart from their alignments, the main evidence that the knots belong
to the same jet (and not to other, smaller outflows) is their
kinematic association with red- and blue-shifted bipolar lobes.
From the radial velocity, the proper motions and the distance 
of the knots from the source it has been
possible to estimate a typical dynamical age of $\sim 10^4$~yr for
these ``giant jets''.

An important characteristic of the giant jets is that they appear
to slow down for increasing distances from the outflow source. This
effect is seen in the HH~34 (Devine et al. 1997) and in the
HH~111 giant jets (Reipurth et al. 1997; Rosado et al. 1999).
The present paper is concerned with the possible theoretical
interpretations of the deceleration effect. This is a 
critical point in the determination of the jet's 
age and also in the identification of the physical properties 
of the central engine that feeds the outflows.

The potential causes that might produce such a deceleration
can be divided into two categories~:
\begin{itemize}
\item internal causes, i.e. that the mechanism is intrinsically related to
the properties of the outflow, for example to a temporal variability of
the ejection velocity,
\item external causes, i.e. that the deceleration is due to the
drag effect resulting from the interaction
of the hypersonic flow with the interstellar medium.
\end{itemize}  

Previous studies have considered both of these possibilities.
Cabrit \& Raga (2000) considered the case of an ejection velocity which
monotonically grows as a function of time, and tried to fit the
observed, position-dependent jet velocity with different, parameterized
forms of the ejection velocity time-dependence.
These authors concluded that the only way to fit the observed
kinematics of the HH~34 giant jet is with an ejection velocity
that slowly increases over $\sim {5\times 10^{4}}$~yr, followed
by a very strongly increasing ejection velocity over the last $\sim 10^4$~yr.

Cabrit \& Raga (2000) argued that this very dramatic increase in
the ejection velocity at recent times appeared to be unlikely, and
then studied an alternative scenario.
Following the idea proposed by Devine et al. (1997), they considered
the knots along the HH~34 giant jet as scattered ``bullets'' resulting from
the combination of an ejection velocity variability and a precession
of the jet axis. Raga \& Biro (1993) have carried out a theoretical
study of this kind of flow, obtaining an analytic description
of this ``machine gun jet'' flow and comparing this model with a
numerical simulation of a radiative, 2D ``slab'' jet with a time-dependent
ejection velocity and direction.

One could argue that the jet/counterjet symmetry observed
in the HH~34 giant jet (see Devine et al. 1997) goes against
the ``environmental drag'' scenario for the deceleration of this
object. As there is no reason to suppose that the environments
within which the jet and the counterjet are traveling have
identical densities, one would think that the drag would introduce
asymmetries between the two outflow lobes.
However, the deceleration induced by the environmental drag
is proportional to ${\rho_a}^{1/3}$ (where
$\rho_a$ is the environmental density, see Cabrit \& Raga 2000),
so that these asymmetries might not be so important.

Also, de Gouveia Dal Pino (2001) presented 3D simulations
(done with the Lagrangian SPH method) of HH~34 assuming a sinusoidal
ejection velocity variability of the kind used by
Raga \& Noriega-Crespo (1998) and studying the two cases of 
a pressured and overpressured jet without considering the 
precession contribution. To reproduce the large spatial working
surface structures the author used a half-amplitude of the velocity
modulation of $\approx 100$~km~s$^{-1}$ and a period of $760$ yr. The
author obtained encouraging results showing that a deceleration of the
jet velocity was obtained from the model (lower decelerations
being produced for initially overpressured jets than for pressure
matched jets). She concluded from the
simulations that the deceleration was related to the temporal velocity
variability of the jet at injection and was mainly caused by progressive
momentum transfer sideways into the surrounding medium by the expelled 
gas from the travelling working surfaces. She also found that a steady 
state jet, with similar initial conditions to those of the pulsed jet,
experienced, on the
contrary, an initial acceleration followed by a constant velocity
propagation regime, which was an additional  indication that the primary
source of
deceleration in giant flows could not be attributed to simple breaking of 
the jet head against the external medium. These models were run over a distance
of only $0.3$ pc, well below the size of the giant HH~34 jet
(see Devine et al. 1997).

In the present paper, we carry out a numerical study of the
effect of a precession of the outflow axis on the deceleration
of a giant jet. The existence of such a phenomenon is suggested by the 
observed morphology of the flow, which shows evidence of 
long period (of the order of $10^4$ yr) precession 
(Devine et al. 1997).
The precession is generally ascribed to tidal forces produced by a companion
in a binary or multiple system. Even though in the
case of HH~34 the binary source has noy yet resolved, there are elements that
indicate that this source could be a binary such as the discovery of a second 
outflow (HH~534) emanating from the source as well as the abrupt change of 
direction of the jet axis near the knots B (Reipurth et al. 2002). 
Less evident is the explanation of the existence of a precession period
of the order of $10^{4}$ yr.

As reported in Terquem et al. (1999),
the precession period ($\tau_{p}$) depends on the orbital parameters and
the spatial extent of the accretion disk. The precession period
is generally at least one order of magnitude larger than the orbital period.
Therefore, to justify such a large $\tau_{p}$ we need a value of the
ratio between the accretion radius disk $R$ and the orbital radius
$r_{0}$ ($\sigma$ $=$ $R$/$r_{0}$ ) of the order of $10^{-3}$
(see Masciadri \& Raga, 2002).

Reipurth (2000) has also proposed that perturbations on an accretion
disk due to close passages (i.~e., at perihelion) of a binary companion
in an elliptical orbit could be responsible for producing a time-variability
in the ejection of the outflow. It is of course unclear whether or not
such a mechanism could produce a variability in the ejection velocity
such as the one included in our jet models.

We carry out the simulations of only one of the 
two lobes of HH~34 (with and without precession)
conserving the same geometrical and physical parameters. However,
our numerical simulations could correspond to any of the two
lobes of the HH~34 giant jet. Actually, our simulations are
made to reproduce the morphology of the northern lobe of the
HH~34 outflow, and in order to compare the predicted maps
with the southern lobe it is necessary to carry out a point
reflection of the predicted maps with respect to the position
of the outflow source. This point symmetry of a precessing jet/counterjet
system is clearly seen in the observations of the HH~34 giant outflow
(Devine et al. 1997).

We underline that this study cannot be applied in a simple way
to other giant jets. The two lobes of HH~111, for example, show
quite straight paths. HH~355, on the contrary, seems to precess
with a half-opening angle of $\sim 13^\circ$
and a period of $\sim 1500$-$2000$ yr (Reipurth et al. 2002).
Further work should be carried out to study the properties of these
other giant flows.

We find that the impact of the precession
on the deceleration mechanism is quite considerable,
and that the dynamical age of the jet grows by
$\sim 3000$~yr when including a precession. We also find that
H$\alpha$ intensity maps and position-velocity diagrams obtained
from models with precession reproduce the observations of the
HH~34 giant jet in a qualitatively successful way (an agreement
which is not found for models without precession).

We add that, from a numerical point of view it is not a simple
exercise to reproduce the evolution of outflows over such
a large spatial and temporal extent particularly given 
the small initial radius of the beam. Indeed, in order 
to cover the whole domain of $1.5$ pc,
previously published simulations used 
$r_{j}$ $=$ $10^{16}$ cm (de Gouveia Dal Pino 2001) and
$r_{j}$ $=$ $10^{17}$ cm jet radii (Cabrit \& Raga 2000), 
which are $1$-$2$ order of magnitude greater than the width of the
jet as observed in HST images (Reipurth \& Raga 1999).
One of the goals of our study is to simulate the HH~34 giant
jet over its full extent ($\sim$ $1.5$ pc) using
the correct $r_{j}$ $=$ $3\times10^{15}$ cm initial jet 
radius corresponding to $0\arcsec.4$ at $460$ pc. We underline
that the radius is measured at $\sim 10\arcsec$ distance away 
from the HH~34 source. 

In Section \ref{param}, we describe the parameters
and the ejection velocity time-variability used in our models
of the HH~34 jet. In Section \ref{numer}
the numerical simulations are discussed, and
H$\alpha$ maps and position-velocity diagrams predicted from models
with and without precession are presented and compared with the
corresponding observations of the HH~34 giant jet.
In Section \ref{conc} we summarize the conclusions
of this study.

\section{Parameters for the jet models}
\label{param}

Following the study of HH~34 of Raga \& Noriega-Crespo (1998),
we attempt to reproduce the structure of this jet by assuming the
existence of a sinusoidal ejection velocity time-variability.
As our work is focussed on trying to reproduce the large scale structure
of this outflow, a single mode time-variability is appropriate,
as opposed to the three-mode variability used by Raga \& Noriega-Crespo
(1998) to model the structures close to the source along the southern lobe.

We therefore consider an ejection velocity of the form
\begin{equation}
v(t)=v_{j} + v_{a}\sin{ \omega_{a}t}
\end{equation}
where $v_{j}$ is the average jet velocity, $\omega_{a}=2\pi/\tau_{a}$
is the frequency ($\tau_{a}$ is the period) and
$v_{a}$ is the half-amplitude of the variability law.

We adopt the following values~: $v_j=300$~km~s$^{-1}$,
$v_a=110$~km~s$^{-1}$ and $\tau_a=1010$~yr. These values are
consistent with the long period mode deduced from the kinematics
of the southern lobe of HH~34 by Raga \& Noriega-Crespo (1998). Also,
these parameters give a spatial separation
\begin{equation}
\Delta x\approx \tau_a v_j={9.5\times 10^{17}}\,{\rm yr}\,,
\label{dx}
\end{equation}
which corresponds to a separation of $\approx 120''$ on the plane
of the sky (considering a distance of 460~pc to HH~34 and a
$\phi=28^\circ$ angle between the outflow axis and the plane of
the sky). This separation is consistent with the separations between
successive knots along the HH~34 giant jet
(see Devine et al. 1997).
Also, the images of the HH~34 giant jet show evidence of a long
period precession. From the images of Bally \& Devine (1994) and
Devine et al. (1997), we find that both the jet and the counterjet
appear to lie within a cone of half-opening angle $\alpha=6^\circ$,
and that the locus of the jet appears to imply a $\tau_p=12000$~yr
precession period. We take these values as estimates of a possible
precession of the HH~34 outflow.

For our numerical simulations, we consider that the jet has a top-hat
initial cross section, of radius $r_j={3\times 10^{15}}$~cm, corresponding
to $0''.4$ at the distance of HH~34, which is consistent with
the recent results obtained from HST images of this object (Reipurth et
al. 2002). Moreover, we choose two initial jet number
densities, $n_j=10^3$ and ${5\times 10^3}$~cm$^{-3}$. We have assumed
that the surrounding environment is homogeneous, with a
$n_a=100$~cm$^{-3}$ number density. We have set the initial temperature
of both the jet and the environment to 1000~K, and assumed that the gas
is neutral, with the exception of Carbon, which is singly ionized.

\section{Numerical simulations}
\label{numer}

\subsection{The numerical computations}
\label{comp}

The simulations were carried out using 3D gasdynamic adaptive
grid yguazu-\'a code. This code integrates the 3D gasdynamic
equations and a set of atomic/ionic reaction equations
for the species HI, HII, OI, OII, OIII, CII, CIII and CIV.
They are the main contributors to the cooling function.
The yguazu-\'a code employs the {\it flux-vector splitting} algorithm
of Van Leer (1982) and it was described in detail by Raga et al. (2000).
The reaction rates and the non-equilibrium cooling function that we
have used are given by Raga et al. (2002a).

The computations were carried out on a 7-level, binary adaptive grid
with a maximum resolution along the three axes of $1.95\times 10^{15}$~cm.
The computational domain extends over $10^{18}\times 10^{18}\times
(4$x$10^{18})$~cm, corresponding to $512\times 512\times 2048$ grid points
at the highest resolution grid level. The jet is injected at the origin
in the centre of the $xy$-plane and the outflow axis precesses around
the $z$-axis.

The maximum resolution (level $7$) is allowed only within a
region limited within a spherical surface of radius $0.5\times 10^{18}$~cm
centred on the injection point. The next highest resolution
(level $6$) is allowed only in a region with an outer radius
of $1.5\times 10^{18}$~cm.

Figure \ref{grid} shows the adaptive grid configuration
on the $xz$-plane obtained from Model B (see below)
after a $t=8.4\times 10^{3}$~s time integration. This figure shows
the grid in a domain of size ($10^{18}$, $3\times 10^{18}$)~cm
(left frame), and two successive zooms (centre and right) showing
the higher resolution grids.

We computed three different models~:
\begin{itemize}
\item{\bf Model A - } a jet that precesses in a cone of
half-angle $\alpha=6^{\circ}$ with a period $\tau_{p}=12000$~yr.
The initial jet density has a $n_{j}=10^{3}$~cm$^{-3}$ value,
resulting in a $\rho_{j}/\rho_{a}=10$ jet to environment density ratio.
\item{\bf Model B - } the same precession as Model A, but a jet
with $n_{j}=5\times 10^{3}$~cm$^{-3}$ ($\rho_{j}/\rho_{a}$ $=$ $50$),
\item{\bf Model C - } a jet with the same parameters as Model A but 
without precession.
\end{itemize}
The three models have the time-dependent ejection velocity
given by Eq. (1), and the jet and ambient medium parameters given at
the end of \S 2.

As an example of the flows that result from our simulations,
in Figure \ref{colden} we show a time-sequence of
the column density obtained for Model B (which, as we show below
is the model that more closely resembles de HH~34 giant jet).
The column density was obtained by integrating the density field along
the $y$-axis. Figure \ref{colden} shows that many working surfaces
are formed, and that they travel in different directions away from
the source as more or less independent ``bullets''. The qualitative
features of this kind of flow were discussed by Raga \& Biro (1993).

\subsection{H$\alpha$ maps}
\label{alfa}
All the three models (A, B and C) were run over different
time intervals such that the outflows have traveled a distance of
$\approx 3\times 10^{18}$~cm, equivalent to $\approx 1$~pc.
Figures \ref{mapaha_evol_a}, \ref{mapaha_evol_b} and
\ref{mapaha_evol_c} show the temporal evolution of the H$\alpha$ maps
predicted from the three models. These maps were computed assuming an
$\phi=28^\circ$ angle between the outflow axis and the plane of the sky.
The H$\alpha$ emission coefficient was calculated considering the contributions
of the recombination cascade and of collisional excitations from the
ground state.

It is evident that the locci of the jets are quite different
for Models A, B and C. Model A and B show a bending not present in Model C.
Moreover the first two models have a larger concentration of knots 
than the Model C. The non-precessing model C, of course,
shows a structure of aligned knots that looks dramatically different from
Models A and B. Also, the times $t_A$, $t_B$ and $t_C$ (corresponding
to Models A, B and C, respectively) at which the jet heads reach a
distance of $1$~pc from the source are substantially different from each other.
We obtain $t_A=18.4\times 10^{3}$~yr, $t_B=8.4\times 10^{3}$~yr and
$t_c=5.4\times 10^{3}$~yr. We note that the maximum H$\alpha$ map value
is equal to $7\times10^{-6}$ erg cm$^{-2}$ s$^{-1}$ sr$^{-1}$
in Model A, equal to $2.5\times10^{-5}$ erg cm$^{-2}$ s$^{-1}$ sr$^{-1}$
in Model B and equal to $2\times10^{-7}$ erg cm$^{-2}$ s$^{-1}$ sr$^{-1}$
in Model C. Comparing the results of our
simulations with the measured H$\alpha$ emission (Reipurth et al. 2002,
Fig. 4) we conclude that the model that better reproduces the observed
H$\alpha$ fluxes is Model B. Indeed, the highest H$\alpha$
contour obtained displayed in the HST images of Reipurth et al.
(2002, Fig.~4) corresponds to $\approx$ [$13$-$18$]
$\times 10^{-18}$ erg cm$^{-2}$ s$^{-1}$
pixel$^{-1}$, which is equivalent to [$3$-$5$] $\times 10^{-4}$
erg cm$^{-2}$ s$^{-1}$ sr$^{-1}$. We conclude that the model
that better reproduces the observed H$\alpha$ fluxes is Model B.

However, we find that the model predicions give H$\alpha$ maps
that are  in general one order of magnitude fainter than
the the observed ones. In a previous study, Raga \&
Noriega-Crespo (1998) found that for an initial jet density
of $5 \times 10^{2}$ $cm^{-3}$, the predicted flux is two orders
of magnitude fainter than HH~34. The fact that we obtain
a one order of magnitude higher flux is consistent with
the fact that in the present model we have considered a
$5 \times 10^{3}$ cm$^{-3}$ jet density (in other words,
the models follow the standard intensity $\propto$ pre-shock
density scaling law). Therefore, in order
to obtain a better agreement with the H$\alpha$ fluxes of HH~34,
we would need a model in which both the jet and the environment
are denser by a factor of $\sim 10$.
We have not computed denser models, because they have shorter
cooling distances, which would not be resolved appropriately in
our numerical simulations.

At the same time, the morphology of the knot distribution in Model B
better resembles the observed morphology of the HH~34 giant
jet (Devine et al. 1997, Fig.~6) than do the knot structures
predicted from Models A and C.
We note that Model A (with $\rho_{j}/\rho_{a}=10$) produces, at
a time $t\approx 11.2\times 10^{3}$~yr from the beginning of the simulation,
an extended shock structure (resulting from a piling up of several
working surfaces) at about $z\approx 0.7\times 10^{18}$~cm.
Such piling ups of working surfaces do not occur in Model B
(which has $\rho_{j}/\rho_{a}=50$), resulting in quite dramatically
different H$\alpha$ maps being predicted from both models.
We therefore conclude that the jet-to-environment density
ratio $\rho_{j}/\rho_{a}$ is a critical parameter for jets from
precessing sources, and has to be adjusted in order to be able
to reproduce the observations of a given HH jet.

The calculation of the temporal evolution of the H$\alpha$
maps permitted us to retrieve the proper motions of the knots
dispersed along the jet path. Figure \ref{mapaha_tot2}
shows H$\alpha$ maps obtained at two different times. From the
positions of the knots in these two frames, we have computed
proper motion vectors (shown on the left plot).
One can observe that the magnitude
of the proper motion velocity slows down as a function of
distance from the source. This result is qualitatively consistent
with the proper motion measurements of Devine et al. (1997), which 
show tangential velocities which decrease
from $198$ $km/sec$ to $96$ $km/sec$ as a function of distance 
from the source along the giant jet.
We do not attempt to carry out a quantitative comparison of the
predicted and observed proper motions, as the errors of the
proper motions of the HH~34 giant jet are quite large.

In order to carry out a more quantitative comparison between
the kinematics of HH~34 and our models, it is better to compare
the  predicted and measured radial velocities. This
kind of comparison between models and observations is
described in the following section.

\subsection{Position-velocity diagrams and radial velocities}
\label{pv}

One of the most reliable ways to validate the simulations
is to compare the PV diagrams obtained from the three models
with observations. Indeed, one of the most interesting
results of Devine et al. (1997, Fig.~7) are their quite accurate
measurements of decreasing radial velocities as a function of
distance from the source along the HH~34 giant jet.
Figure \ref{diag_pv} shows the PV diagrams predicted  obtained from 
Models A, B and C.
The black, thin line represents the maximum of the emission (as
determined from quadratic fits to the line peaks)
vs. distance from the source. The three PV diagrams
are calculated at the times at which the head of the jet
reaches $3\times 10^{18}$~cm (these times are $t_A$, $t_B$
and $t_C$, see \S \ref{alfa}).
The same $\phi=28^\circ$ angle between the outflow axis and the plane
of the sky was considered.

Figure \ref{mapaha_tot1} shows the radial 
velocity vs. distance from the source measured
by Devine at al. (1997) for the knots
along both the north and south lobes of the HH~34 giant jet.
We changed the sign of the velocities of the knots 
in the south lobe, so that all of the velocities are then positive.
The bold points are the peak of the gaussian fit and the
error bars represent the FWHM of the emission line. 
We omitted the point related to the HH~34~X knot because, as one can see in 
Figure 7 of Devine et al. (1997), it is quite distant 
from the trend defined by the velocities of all of the other knots
in the outflow.

The three continuous lines join the radial velocities of the working
surfaces (the sharp velocity jumps in Fig. \ref{diag_pv})
obtained from Model A (dotted line), Model B (thin line) and Model C
(dot-dash-dot line). These radial velocities (shown with
stars in the PV diagrams of Fig. \ref{diag_pv}) correspond to the
successive knots which are seen in the predicted H$\alpha$ maps.
The positions of the stars correspond to the intensity maxima
(i.~e., the knots) in the corresponding H$\alpha$ maps.

Both Figures \ref{diag_pv} and \ref{mapaha_tot1}
show that Model A produces too steep radial velocity vs. distance
decrease, particularly beyond $2\times 10^{18}$~cm from the source.
Model C produces radial velocities which are too high over the whole
path of the jet. On the other hand, Model B produces radial
velocities which agree quite well with the observed values.

Figure \ref{mapaha_tot1}
(right-hand side) shows linear fits to the
radial velocities as a function distance from the source obtained from
the observations and from
Models A, B and C, in other words, the lines trace the
decreasing rate of the radial velocity. In this graph, we
again see that while Model A produces a too sharp drop in radial
velocities, and Model C gives velocities which are too high,
Model B does produce a good agreement with the observations.

\subsection{Effects of the precession vs. the initial jet to
environmental pressure ratio}

What can we say about the impact of the precession and
of the initial jet to environmental pressure ratio ? 
We define $\eta$ as the ratio between the initial jet and 
environmental density ($\eta$ $=$ $\rho_{j}/\rho_{a}$) and we define
$k$ as the corresponding pressure ratio ($k$ $=$ $p_{j}/p_{a}$). 
Knowing the values
of $\rho_{a}$, $\rho_{j}$, $T_{a}$ and $T_{j}$ we can calculate
$k$ $=$ $10$ for the Models A and C and
$k$ $=$ $50$ for Model B. All three models
are therefore overpressured, Model A and C have the same
value of $k$ and Model B is more highly overpressured. 
Figure \ref{mapaha_tot1} shows that the
deceleration rates of Models A and C are very
different in spite of the fact that the two models have 
identical $k$. Considering that in Model A we
have a precession that is absent in the Model C we would
then conclude that the difference in the deceleration 
 is a direct result of the precession.\newline 
de Gouveia Dal Pino (2001) studied the effects of different 
$k$ values on a non-precessing jet and found that lower 
values of $k$ resulted in stronger decelerations for the working
surfaces. This is a result of the fact that the different $k$
values are obtained by changing the environmental 
temperature $T_{a}$, with higher values of $T_{a}$ resulting
in a stronger coupling between the aligned working surfaces
and the surrounding gas (and therefore causing a stronger deceleration).
This rather subtle effect is not likely to be important in the case
of a precessing jet for which the scattered working surfaces ram directly
into the undisturbed environment. For this case, the important 
parameter for determining the deceleration rate is the density ratio $\eta$.

It would be interesting to study in the future 
the effect on the deceleration mechanism of two other flow parameters: 
the angle and the period of the precession. However, an idea of
the effects that would be introduced can be obtained from the analytic
model of Raga \& Biro (1993). These authors showed that the distance
(measured along the precession axis) at which the working surfaces
become independent ``bullets'' interacting with the surrounding
environment is $z_b\approx r_j \tau_p\cot\alpha / (\pi \tau_a)$,
corresponding to $\approx {4\times 10^{17}}$~cm for our model B.
Therefore, in our model, the $z<z_b$ region in which the working
surfaces are sheltered from a direct interaction with the environment
is small, so that they are subjected to the full environmental drag
during most of their evolution. We would then expect that all models
with $\alpha$, $\tau_p$ and $\tau_a$ such that $z_b$
is much smaller than the full length of the jet will have
deceleration properties similar to our models, provided that the
they have the same $r_j$ and jet-to-environment density ratios.

\section{Conclusion}
\label{conc}

In this paper, we explore the role of a precession of the outflow
axis on the deceleration of giant HH jets. In particular, we try to
simulate the HH~34 giant jet, and we compare the results of our
numerical simulations with previously published observations of this
object.

Our simulations represent a step forward from previous numerical
studies of HH objects, in that the use of a 7-level, binary
adaptive grid has allowed us to compute models which appropriately
cover all of the relevant scales of a giant jet, from the
$\sim 100$~AU jet radius close to the source to the $\sim 1$~pc
length of the outflow. Previous simulations of giant jets either
did not cover the length of a real flow (de Gouveia dal Pino 2001),
or else had a very large jet radius (Cabrit \& Raga 2000 and 
de Gouveia dal Pino 2001).

A set of simulations done with and without precession of the
outflow axis are presented, and predictions of
H$\alpha$ maps, proper motions and radial velocities are compared
with the observations of Devine et al. (1997).
The principal conclusions of our study are the following~:
\begin{itemize}
\item we see that the morphology and kinematics of the HH~34
giant jet can be reproduced with a model of a jet with
a sinusoidal ejection velocity variability (with a mean
velocity of 300~km~s$^{-1}$, a half-amplitude
of 100~km~s$^{-1}$ and a period of 1010~yr) and a precession
of the outflow axis (with a half-angle of 6$^\circ$
and a 12000~yr period).
The simulated and measured H$\alpha$ maps and radial
velocities show a good qualitative as well as quantitative agreement,
\item comparing simulations done with and without precession 
we showed that the simple precession can give differences in the 
jet age estimations of the order of about $3000$ yr. This proves that the 
drag effect produced by the external medium on the working surfaces 
is not negligible with respect to the $\sim 10^{4}$ yr dynamical
timescale of the outflow.
\item we proved that the $\rho_{j}/\rho_{a}$ ratio is a
critical parameter in the determination of the deceleration
rate of the jet, and that it has to be properly adjusted in
order to be able to fit the observed properties of a giant
HH flow.
\end{itemize}

The results of our study do not exclude that, in other giant jets,
a correct jet deceleration could be attained without the precession. 
More models should be tested with different 
$\rho_{j}/\rho_{a}$ ratios to obtain more definite conclusions. 
Besides this, in the case of HH~34, the results seem to indicate that 
the precession has a fundamental role in the deceleration mechanism.
We underline that the non-precessing, velocity-variable jet of Model C
also decelerates, however it has radial velocities which are larger than
the observed ones in HH34 (Devine et al. 1997). Previous 3D modeling of
time-variable, non-precessing giant outflows (de Gouveia Dal Pino
2001) had also detected 
jet deceleration that reproduced the observations only qualitatively. 
Therefore, the results of the present work indicate that
in the case of HH34, it is the combined effect of both, the jet temporal
velocity variability and the precession (along with the appropriate choice
of the ratio $\rho_j/\rho_a$) that reproduces the observed deceleration
pattern in that source (as in Model B). One could argue that
Model C with different
$\rho_{j}/\rho_{a}$ ratios or velocity variability law could reproduce the 
correct deceleration. On the other hand we observe that a change of the 
$\rho_{j}/\rho_{a}$ ratio seems to produce a modification in the deceleration rate
(Fig.~\ref{mapaha_tot1} right hand side). A different velocity variability 
would produce a different distribution of the working surfaces along the
jet trajectory and an H${\alpha}$ map characterized by a different emission. 
We also recall that the emission of the H${\alpha}$ maps strongly depends on the $\rho_{j}/\rho_{a}$ ratio. 
A smaller $\rho_{j}/\rho_{a}$ ratio in Model C would probably reproduce a H${\alpha}$ map characterized by an emission level which is too low (see Section \ref{alfa}).
It therefore appears that these 
high radial velocities are due neither to an incorrect $\rho_{j}/\rho_{a}$ 
ratio nor to a velocity variability law different from the one that we have considered.

In our simulations of the HH~34 giant jet, all of the structure
of the outflow is due to a velocity variability with a single,
sinusoidal mode and a well ordered precession. In our model,
both the velocity variability and the precession last for
all of the life of the outflow. Furthermore, the ejection velocity
variability that we have used agrees with the one determined
by Raga \& Noriega-Crespo (1998) for the region between the
source and HH~34S, so that there is evidence that this
variability is continuing to at least quite close to
the present time.

The existence of a precession is an indication that the source
belongs to a binary or multiple system. This is in agreement
with Reipurth (2000), who argued that the sources of giant
HH jets are binary or multiple systems.

However, our results seem to be less consistent with the thesis
proposed by Reipurth (2000) that
giant jets are fossil records of the evolution
of orbital motions in disintregating multiple systems. This
process has three distinct phases~:
(1) a non-hierarchical state
(called {\it interplay}) in which the multiple system performs
a random motion, (2) a close triple approach in which a
close binary is formed and a low mass star or embryo moves
over to a larger orbit and (3) an ejection phase in which the
low mass embryo is ejected from the nucleus of the system.

In our models, the structure of the HH~34 giant jet is reproduced
without the need of having different properties of the ejection
at different times, reflecting qualitative changes in the outflow
source resulting from the three phases of a disintegrating multiple
system (see above). Therefore, we conclude that if the HH~34 system
does correspond to an outflow history with distinct phases, the evidence
for this appears to have been lost in the complexities of the
interaction between the jet and the surrounding environment.

\acknowledgments
The work of EM and AR was supported by CONACyT grants 34566-E and 36572-E.
We thank Israel D\'\i az for installing the new computer with which
the numerical simulations were carried out.
E.M.G.D.P. has been partially supported by grants of the Brazilian
Agencies FAPESP (grant 97/13084-3) and CNPq. A.N-C. research was 
carried at the Jet Propulsion Laboratory, California Institue of Technology, 
under contract with NASA; and partially funded by ADP grant NRA0001-ADP-096.
We acknowledge an anonymous referee for giving us helpful comments.

\newpage

\begin{figure*}
\centering
\figcaption{7 level, binary adaptive grid obtained from Model B after
a $8.4\times10^{3}$ yr time integration. The abscissas and ordinates
are labeled in units of $10^{18}$~cm. On the left hand side,
the grid extended over the ($10^{18}$, $3\times10^{18}$) cm domain
is shown. In the centre and right hand side, successive zooms of
smaller regions are shown. The points represent the central positions
of the grid cells on the xz-plane.
\label{grid}} 
\end{figure*} 

\begin{figure*}
\centering
\figcaption{Temporal evolution of the column density (integrated along
the $y$-axis) obtained from Model B. The abscissas and ordinate
are labeled in units of $10^{18}$~cm. The column densities are shown
with the linear grey-scale given (in cm$^{-2}$) by the bar on the right.
\label{colden}}
\end{figure*}

\begin{figure*}
\centering
\figcaption{Temporal evolution of the H$\alpha$ maps obtained
from {\bf Model A}. The last frame corresponds to the time at which
the jet extends out to $\approx$ $3\times10^{18}$ cm from the source.
The abscissas and ordinates are labeled in units of $10^{18}$~cm.
The H$\alpha$ maps are shown with the logarithmic grey-scale given
(in erg cm$^{-2}$ s$^{-1}$ sr$^{-1}$) by the bar on the right.
The maximum H$\alpha$ intensity value is of $7\times10^{-6}$
erg cm$^{-2}$ s$^{-1}$ sr$^{-1}$. 
\label{mapaha_evol_a}}
\end{figure*}

\begin{figure*}
\centering
\figcaption{Temporal evolution of the H$\alpha$ maps obtained from
{\bf Model B}. The last frame corresponds to the time at which the
jet extends out to $\approx$ $3\times10^{18}$ cm from the source.
The abscissas and ordinates are labeled in units of $10^{18}$~cm.
The H$\alpha$ maps are shown with the logarithmic grey-scale given
(in erg cm$^{-2}$ s$^{-1}$ sr$^{-1}$) by the bar on the right.
The maximum H$\alpha$ intensity is of
$2.5\times10^{-5}$ erg cm$^{-2}$ s$^{-1}$ sr$^{-1}$.
\label{mapaha_evol_b}}
\end{figure*}

\begin{figure*}
\centering
\figcaption{Temporal evolution of the H$\alpha$ maps obtained from
{\bf Model C}. The last frame corresponds to the time at which the
jet extends out to $\approx$ $3\times10^{18}$ cm from the source.
The abscissas and ordinates are labeled in units of $10^{18}$~cm.
The H$\alpha$ maps are shown with the logarithmic grey-scale given
(in erg cm$^{-2}$ s$^{-1}$ sr$^{-1}$) by the bar on the right.
The maximum H$\alpha$ intensity is of
$2\times10^{-7}$ erg cm$^{-2}$ s$^{-1}$ sr$^{-1}$.
\label{mapaha_evol_c}}
\end{figure*}

\begin{figure}
\centering
\figcaption{The left plot shows the H$\alpha$ map at $6\times10^{3}$ yr.
The right hand side plot shows the H$\alpha$ map at $6.4\times10^{3}$ yr.
On the left plot the proper motions of the four brightest knots are marked.
The abscissas and ordinates are labeled in units of $10^{18}$~cm.
The H$\alpha$ maps are shown with the logarithmic grey-scale given
(in erg cm$^{-2}$ s$^{-1}$ sr$^{-1}$) by the bar on the left.
\label{mapaha_tot2}} 
\end{figure}

\begin{figure*}
\centering
\figcaption{Position velocity diagram obtained for the Models A
(left hand side), B (centre) and C (right hand side). The emission
is shown with a logarithmic greyscale given (in erg~s$^{-1}$~cm$^{-1}$
(cm/s)$^{-1}$~sr$^{-1}$).
The black,
thin line represents the emission maximum of the PV diagram vs.
distance from the source. On the abscissas, the radial velocity values
are given in 100~km/s units. The ordinates give the position along
the spectrograph slit in $10^{18}$~cm units. The abscissas give 
the absolute values of the radial velocities.
\label{diag_pv}} 
\end{figure*}

\begin{figure*}
\centering
\figcaption{On the left hand side are shown the absolute values of the 
radial velocity vs.
distance from the source measured by Devine et al. (1997) (bold points
and errors bar). The graph also shows the radial velocities obtained
from Model A (dotted line), Model B (thin line) and Model C
(dot-dash-dot line). On the right hand side, linear fits to the
values reported in the left plot are shown.
\label{mapaha_tot1}} 
\end{figure*}

\end{document}